\documentstyle[preprint,pra,aps,epsf,tighten]{revtex}

\begin{document}
\draft
\title{Resonant processes in a frozen gas }
\author{J. S. Frasier, V. Celli and T. Blum}
\address{Department of Physics, University of Virginia, \\
Charlottesville, VA 22901.}
\date{\today}
\maketitle

\begin{abstract}
We present a theory of resonant processes in a frozen gas of atoms
interacting via dipole-dipole potentials that vary as $r^{-3}$,
where $r$ is the interatomic separation. We supply an exact result
for a single atom in a given state interacting resonantly with a
random gas of atoms in a different state. The time development of
the transition process is calculated both on- and off-resonance,
and the linewidth with respect to detuning is obtained as a
function of time $t$. We introduce a random spin Hamiltonian to
model a dense system of resonators and show how it reduces to the
previous model in the limit of a sparse system. We derive
approximate equations for the average effective spin, and we use
them to model the behavior seen in the experiments of Anderson
\emph{et al.}\cite{anderson} and Lowell \emph{et
al.}\cite{lowell,lowell2}. The approach to equilibrium is found to
be proportional to $\exp (-\sqrt{\gamma _{eq}t}$ ), where the
constant $\gamma _{eq}$ is explicitly related to the system's
parameters.
\end{abstract}





\narrowtext


\section{Introduction}

\label{intro}Frozen gases are a new and in some ways ideal
laboratory to test our understanding of quantum theory in a
complex system. With present technology one can manipulate and
detect electronic processes with an extraordinary selectivity and
precision; furthermore, the translational temperature of the gas
can be lowered to the point where it can be ignored when
discussing electronic processes. Frozen Rydberg gases have the
added advantages that their states are well understood and many
processes occur on a microsecond time scale, easily allowing for
time-resolved spectroscopy. Pioneering experiments on resonant
processes in these gases have been carried out by Anderson
\emph{et al.}\cite{anderson} and Mourachko \emph{et al.}\cite
{mourachko}. The present work was motivated by the desire to
understand those experiments and the subsequent work of Lowell
\emph{et al.}\cite{lowell,lowell2} highlighting the dynamic
aspects of these resonant, many particle systems. The measurements
are of mixtures of ${}^{85}Rb$ atoms initially prepared in the
$23s$ and $33s$ states (henceforth to be called the $s$ and
$s^{\prime }$ states, respectively). There are initially $N$ atoms
in the $s$ state and $N'$ atoms in the $s'$ state.
The transition $ss^{\prime }\rightarrow pp^{\prime }$ is monitored, where $%
p $ refers to the $24p$ state and $p^{\prime }$ refers to the
$34p$ state, at and near resonance, where $\epsilon _{s}^{\prime
}+\epsilon _{s}=\epsilon _{p}^{\prime }+\epsilon _{p}$. Some of
the experimental features we wish to understand are the rapid rise
followed by slow approach to saturation of the signal and the
width of the lineshape (in the detuning $\Delta =\epsilon
_{p}^{\prime }+\epsilon _{p}-\epsilon _{s}^{\prime }-\epsilon
_{s}$) as a function of time.

Each individual interaction leads to a (coherent) oscillatory
behavior, but as we will see below, averaging the $ss^{\prime
}\rightarrow pp^{\prime }$ interaction over the random positions
of the atoms greatly smooths out the signal. The effective
``incoherence'' brought about by the $sp\rightarrow ps$ process
(the ``walking away'' discussed by Mourachko \emph{et
al.}\cite{mourachko}) completes the smoothing out of the
on-resonance signal but has less effect on the off-resonance
signal. (The $s^{\prime }p^{\prime }\rightarrow
p^{\prime }s^{\prime }$ process is also present, but is smaller than $%
sp\rightarrow ps$ by a factor of $16$ in this case and can
therefore be neglected to the first approximation.)

It was surmised by Anderson \emph{et al.}\cite{anderson} that the
linewidth must be of
the order of the average interaction energy. This interaction splits the $%
ss^{\prime }$-$pp^{\prime }$ degeneracy and the subsequent migration of the $%
p$ state to other atoms broadens the energy of this ``elementary
excitation'' into a band, as appropriate to an amorphous solid.
Cluster calculations were performed to illustrate this band
formation\cite{lowell}, but averaging over atom positions was done
by order-of-magnitude arguments only. Mourachko \emph{et
al.}\cite{mourachko} pointed out, in connection with their
experiments on a different system, that the ``walking away'' of
the $p$ excitation from its original location should be regarded
as a diffusion process, and dynamical equations for the resonance
transition in the presence of this diffusion were written
down\cite{akulin}.

The present work builds on these earlier insights, but goes
considerably further in the process of accurately averaging over
atomic positions. It turns out that this averaging (effectively, a
phase averaging) itself produces an $\exp (-\sqrt{\gamma _{eq}t})$
dependence at large times, characteristic of a diffusive process.
A report of the newer experimental data by Lowell \emph{et al.} in
comparison with the results of the present theory is being
submitted separately\cite{lowell2}. In this paper we give a full
account of the general theory, and we discuss in detail in
Section~\ref{dilute} a particular case that cleanly shows the
effects of randomness and phase-averaging. In this simple case,
the averaging process can be carried out exactly. For the
practically important $r^{-3}$ interaction potential, one can
proceed in a mathematically elegant way and obtain closed formulas
that can be evaluated by a computer package, or in some cases by
hand. It appears to be a fortunate coincidence that the $r^{-3}$
potential accurately describes the dipole-dipole interactions that
are so prevalent in nature. Some aspects of this approach are
easily extended to more general interactions; however, the
$r^{-3}$ dependence has special properties which, when coupled
with the assumption of a random distribution of atoms, lead to
relatively simple results.

Although this work was stimulated by experiments on Rydberg atoms,
it is generally applicable to resonances induced by dipole-dipole
interactions. These may be common in highly excited gases and in
molecular systems. Frozen, resonant gases may even exist in
interstellar clouds.

We develop the theory in three stages. First, in
Section~\ref{dilute}, we discuss a single $s^{\prime }$ atom
interacting resonantly with a surrounding gas of $s$ atoms,
without any possibility for the $s$ excitation to ``walk away''
through an $sp\rightarrow ps$ process. When the result is averaged
over the positions of the $s$ atoms, it corresponds to a
``sparse'' system of $s'$ atoms, i.e. a system where $N'\ll N$. We
do not know of an existing experiment to which this treatment
applies, so this section can be viewed as a theoretical prediction
of the outcome of a possible experiment. Our purpose here is also
one of exposition: this example allows us to introduce in the
simplest context some of the mathematical techniques used
throughout the paper. In Section~\ref {decay}, we complicate
things somewhat by allowing the $s$ state to decay away
exponentially (or more generally, but explicit formulas are given
for the exponential decay). If the $s$ state decays diffusively,
our treatment here is related to that of Mourachko \emph{et
al.}\cite{mourachko}, except for the introduction of
mathematically exact phase averaging. Here, too, the treatment
applies only to the case where $N'\ll N$. Finally, in
Section~\ref{spin} we introduce an effective spin Hamiltonian
which fully models the atomic system for all values of $N$ and
$N'$, as well as for any strength of the resonant process $\left(
ss' \rightarrow pp'\right) $ and of the mixing processes ($sp
\rightarrow ps$ and $s'p' \rightarrow p's'$). This system is
relevant to the experiments of Anderson \emph{et al.}
\cite{anderson} and Lowell \emph{et al.}\cite{lowell,lowell2}. We
show its equivalence to the previous results in the sparse limit
$\left( N'\ll N \right) $ and discuss briefly the approximate
solution in the non-sparse case. The results in this section are
summarized elsewhere\cite{lowell2}, but the derivation is
presented here for the first time, as are additional formulas. The
limitations of the present theory and some of the many possible
extensions are discussed briefly in Section~\ref{discuss}.

\section{Sparse $ss^{\prime }\rightarrow pp^{\prime }$ processes}

\label{dilute}

We consider one atom, initially in the state $s^{\prime }$, in
interaction with a gas of $s$ atoms through an $ss^{\prime
}\rightarrow pp^{\prime }$ process. Like Mourachko \emph{et
al.}\cite{mourachko}, we describe the system by the equations:
\newcounter{tlet}
\setcounter{tlet}{1}
\renewcommand{\theequation}{\arabic{equation}\alph{tlet}}
\begin{eqnarray}
\label{dilute-eqns}
 i~\dot{a}_{0} &=&\Delta ~a_{0}+\sum_{k}V_{k}~c_{k}, \\
\stepcounter{tlet}\addtocounter{equation}{-1}\label{a0dot}
 i~\dot{c}_{k} &=&V_{k}~a_{0}.
\label{ckdot}
\end{eqnarray}
\renewcommand{\theequation}{\arabic{equation}}
Here $a_{0}(t)$ is the
amplitude of the state in which the atom at the origin is in state $%
s^{\prime }$ and all other atoms are in state $s$, while
$c_{k}(t)$ (with $k$ running from $1$ to $N$) is the amplitude of
the state in which the atom at the
origin is in state $p^{\prime }$ and the atom at ${\bf r}_{k}$ is in state $%
p $, while all the others remain in state $s$. $V_{k}$ is the
interaction potential and
$\Delta=\epsilon_{p}'+\epsilon_{p}-\epsilon_{s}'-\epsilon_{s}$ is
the detuning from resonance. We will assume that $V_{k}$ is of the
dipole-dipole form
\begin{equation}
V_{k}=\frac{\mu \mu ^{\prime }}{r_{k}^{3}}.
\end{equation}
The atoms are sufficiently cold that during the time scale of
interest they move only a very small fraction of their separation,
and therefore $V_{k}$ can be taken to be independent of time. As
the temperature increases one approaches the opposite limit, where
binary collisions control the resonant process\cite{stoneman}.
There must of course be a gradual transition, which is beyond the
scope of this paper.

The set of differential equations (1) can easily be solved with
the initial condition $a_{0}(t=0)=1$ and $c_{k}(t=0)=0$ for all
$k$, yielding
\begin{equation}
c_{k}\left( t\right) =-~\frac{2i~V_{k}~e^{-i\Delta t/2}~\sin \left( \sqrt{%
\Delta ^{2}+4{\cal V}^{2}}\,t/2\right) }{\sqrt{\Delta ^{2}+4{\cal
V}^{2}}}\;, \label{ck}
\end{equation}
with
\begin{equation}
{\cal V}^{2}=\sum_{k=1}^{N}V_{k}^{2}\,.
\end{equation}
Eq.~(\ref{ckdot}) then gives $a_{0}(t)$. The experiment monitors
an optical
transition from the $p^{\prime }$ state (or, equivalently, from one of the $%
p $ states), which is proportional to
\begin{eqnarray}
S(t) &=& 1-\left| a_{0}\left( t\right) \right| ^{2}=\sum_{k}\left|
c_{k}\left( t\right) \right| ^{2} \nonumber \\ &=&
\frac{4{\mathcal{V}}^{2}\sin^{2} \left(
\sqrt{\Delta^{2}+4{\mathcal{V}}^{2}}\,t/2 \right)}
{\Delta^{2}+4{\mathcal{V}}^{2}}. \label{S}
\end{eqnarray}
We see that the signal $S\left( t \right)$ exhibits Rabi
oscillations.

We apply this result to a system of $N^{\prime }$ atoms, initially in state $%
s^{\prime }$, randomly dispersed among a much larger number, $N$,
of $s$ atoms. In this sparse limit, the signal is proportional to
$N^{\prime }$ times the sample average of $S(t)$, and the sample
average is equivalent to an ensemble average over the atomic
positions ${\bf r}_{k}$ that are hidden in ${\cal V}^{2}$ (see
Eq.~(\ref{avexp}) below). In general, averaging is more easily
done on the Laplace transform of the signal, which in this case is
\begin{eqnarray}
\tilde{S}(\alpha ) &=&\int_{0}^{\infty }e^{-\alpha t}\,S(t)\,dt
\nonumber \\ &=& \frac{1}{\alpha}
\frac{2{\mathcal{V}}^{2}}{\alpha^{2} + \Delta^{2} +
4{\mathcal{V}}^{2}} \nonumber \\ &=&\frac{1}{2\alpha }\left(
1-\frac{\alpha ^{2}+\Delta ^{2}}{\alpha ^{2}+\Delta ^{2}+4{\cal
V}^{2}}\right) \label{Stild}
\end{eqnarray}
Later in this paper, more complex models are solved directly by
Laplace (or Fourier) transforms; thus Eq.~(\ref{Stild}) is useful
for comparison with these more general results and approximations.

\subsection{Averaging over atom positions}

To compute ensemble averages, we use the following result, valid
in the limit $N\gg 1$, for a random distribution of the variables
${\bf r}_{k}$:
\begin{eqnarray}
\left( e^{-\beta {\cal V}^{2}}\right) _{av} &=&\hspace{-0.2cm}\frac{1}{%
\Omega ^{N}}\int d^{3}r_{1}\ldots d^{3}r_{N}\,\exp \left[ -\beta
\sum_{k=1}^{N}V^{2}(r_{k})\right]  \nonumber \\ \hspace{-0.2cm}
&=&\hspace{-0.2cm}\left\{ \frac{1}{\Omega }\int d^{3}r\exp \left[
-\beta V^{2}(r)\right] \right\} ^{N}  \nonumber \\ \hspace{-0.2cm}
&=&\hspace{-0.2cm}\left\{ 1-\frac{1}{\Omega }\int d^{3}r\left[
1-e^{-\beta V^{2}(r)}\right] \right\} ^{N}  \nonumber \\
&\longrightarrow &\exp \left\{ -\frac{N}{\Omega }\int
d^{3}r\,\left[ 1-e^{-\beta V^{2}(r)}\right] \right\} ,
\label{avexp}
\end{eqnarray}
where $\Omega $ is the volume of the gas in the trap. In
particular, for the dipolar interaction $V(r)=\mu \mu ^{\prime
}/r^{3}$, we have
\begin{equation}
\int d^{3}r\left[ 1-e^{-\beta \left( \mu \mu ^{\prime }\right)
^{2}/r^{6}}\right] =\frac{4\pi ^{3/2}}{3}~\mu \mu ^{\prime
}~\sqrt{\beta }, \label{aveintsimple}
\end{equation}
leading to
\begin{equation}
\left( e^{-\beta {\cal V}^{2}}\right) _{av}=e^{-v\sqrt{\beta }},
\label{ave}
\end{equation}
where
\begin{equation}
v=\frac{4\pi ^{3/2}}{3}~\frac{N}{\Omega }~\mu \,\mu ^{\prime }.
\label{v}
\end{equation}
For the typical densities in the experiments of Anderson \emph{et
al.}\cite {anderson} and Lowell \emph{et
al.}\cite{lowell,lowell2}, $v$ is on the order of MHz. (In
Appendix~\ref{angular-appendix}, we repeat the arguments above
with an interaction that includes the angular dependence due to
the relative positions of the dipole moments and assumes that all
of the dipole moments point in the same direction.) Since $v$ has
the units of an energy, it must be of order $\mu\mu'N/\Omega $ on
dimensional grounds. It is still remarkable, however, that
Eq.~(\ref{v}) gives simply and exactly the quantity $v$ that will
enter in all the averaged quantities in this section.

Using Eq.~(\ref{ave}), we can evaluate the average of any function $F({\cal V%
}^{2})$ provided that it can be represented as
\begin{equation}
F({\cal V}^{2})=\int_{0}^{\infty }d\beta ~\tilde{F}(\beta )~e^{-\beta {\cal V%
}^{2}}.
\end{equation}
Further, we can use the following identity
\begin{equation}
e^{-v\sqrt{\beta }}=\frac{1}{\sqrt{\pi }}\int_{0}^{\infty }dy\,\exp \left\{ -%
\frac{y^{2}}{4}-\frac{v^{2}\beta }{y^{2}}\right\} ,
\end{equation}
so that
\begin{eqnarray}
F({\cal V}^{2})_{av} &=&\int_{0}^{\infty }d\beta \,\tilde{F}(\beta )\,e^{-v%
\sqrt{\beta }}  \nonumber \\ &=&\frac{1}{\sqrt{\pi
}}\int_{0}^{\infty }dy\,e^{-y^{2}/4}\int_{0}^{\infty }d\beta
\tilde{F}(\beta )e^{-v^{2}\beta /y^{2}}  \nonumber \\
&=&\frac{1}{\sqrt{\pi }}\int_{0}^{\infty }dy\,e^{-y^{2}/4}F\left( \frac{v^{2}%
}{y^{2}}\right) .  \label{trik}
\end{eqnarray}
This last set of relations implies that we can average a function $F({\cal V}%
^{2})$ over the positions of the interacting dipoles (atoms in our
case) by
replacing ${\cal V}^{2}$ with $v^{2}/y^{2}$ and integrating over the kernel $%
\exp \left( -y^{2}/4\right) $ $/\sqrt{\pi }$. This trick is not
always useful, as it can lead to highly oscillatory integrals. It
is, however, convenient in determining the lineshape in the
saturation ($t\rightarrow \infty $) limit (as shown below).

\subsection{The averaged signal and lineshape}

\begin{figure}[tbp]
\begin{center}
\leavevmode \hbox{\epsffile{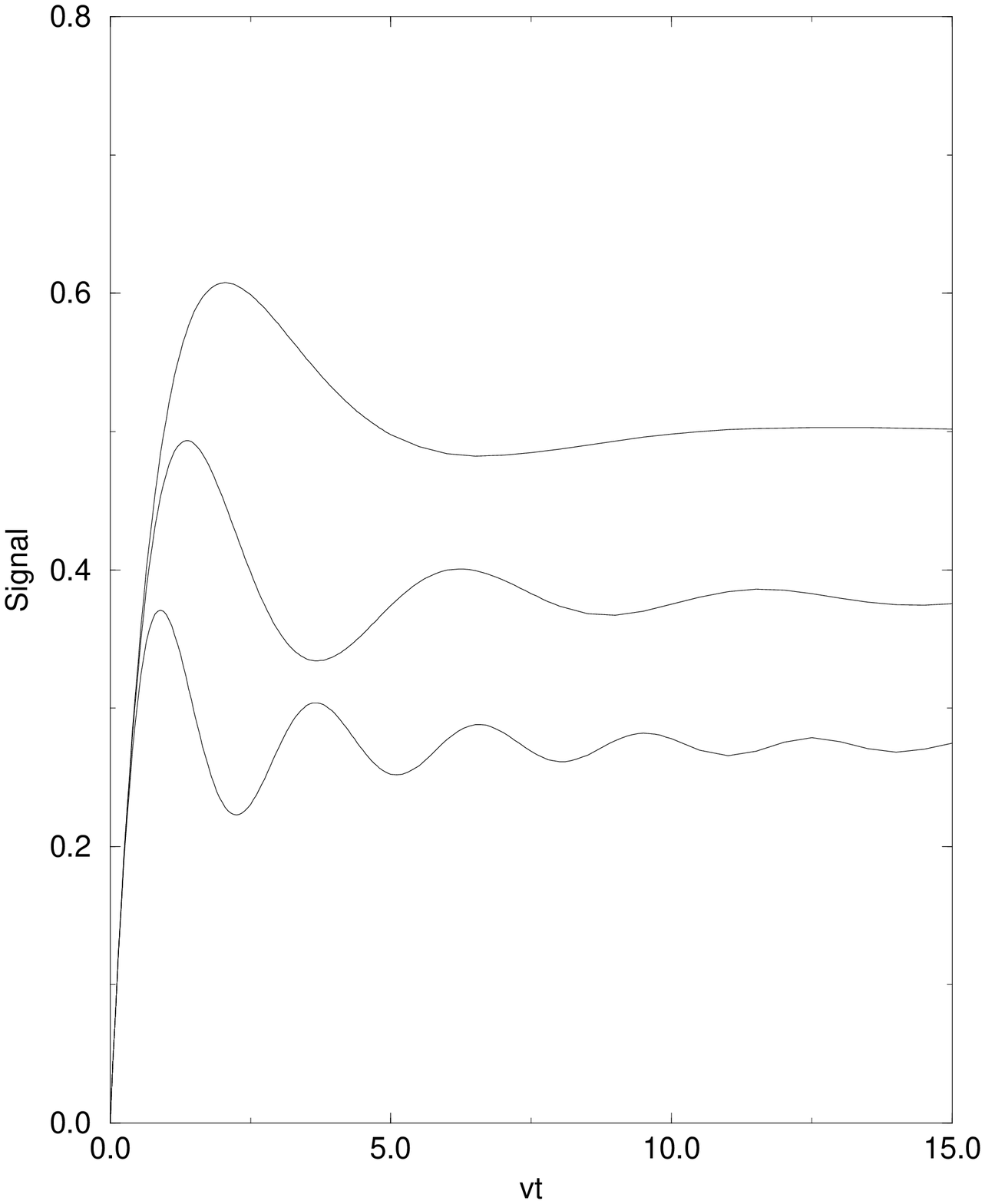}}
\end{center}
\caption{ The averaged
signal, $S(t)_{av}$, for $\Delta/v=0$, $1$, and $2$ (the upper,
middle, and lower curves, respectively) calculated from
Eq.~(\ref{needsanumber}) for the sparse limit. The $ss^{\prime}$-
$pp^{\prime}$ process is averaged; the $sp$-$ps$ process
neglected. The Rabi oscillations are not completely washed out by
the averaging process and are more pronounced off-resonance. }
\label{figure1}
\end{figure}

Starting from Eq.~(\ref{Stild}) in the form
\begin{equation}
\tilde{S}(\alpha )=\frac{1}{2\alpha }-\frac{\alpha ^{2}+\Delta
^{2}}{2\alpha
}\int_{0}^{\infty }d\beta ~e^{-\beta (\alpha ^{2}+\Delta ^{2}+4{\cal V}%
^{2})},
\end{equation}
and using Eq.~(\ref{ave}), we obtain
\begin{eqnarray}
\tilde{S}(\alpha )_{av} &=&\frac{1}{2\alpha }-\frac{A^{2}}{2\alpha }%
\int_{0}^{\infty }d\beta ~e^{-\beta A^{2}-2v\sqrt{\beta }}
\nonumber \\
&=&%
{\displaystyle {\sqrt{\pi } \over 2}}%
\frac{v}{\alpha A}~\exp \left( \frac{v^{2}}{A^{2}}\right) {\rm
erfc}\left( \frac{v}{A}\right) ,  \label{Stld2}
\end{eqnarray}
where $A^{2}=\alpha ^{2}+\Delta ^{2}$. Expanding in $v/A$ and
performing the inverse Laplace transform leads to
\begin{equation}
S(t)_{av}=\frac{\sqrt{\pi }}{2}vt \sum_{n=0}^{\infty }\frac{(-vt)^{n}}{%
\Gamma \left( n+2\right) \Gamma \left( \frac{n+2}{2}\right) }
\ _{1}F_{2}\left[ \frac{n+1}{2};\frac{n+2%
}{2},\frac{n+3}{2};-\frac{\Delta ^{2}t^{2}}{4}\right] ,
\label{hypergeo}
\end{equation}
where $_{n}F_{m}(a_{1},\ldots ,a_{n};b_{1},\ldots ,b_{m};z)$ is a
generalized hypergeometric function. It is known that $_{1}F_{2}$
is related to the integral of a power multiplied by a Bessel
function. Expanding $_{1}F_{2}$ yields the double series
\begin{equation}
S(t)_{av}=\frac{\sqrt{\pi }}{2}vt\sum_{n=0}^{\infty }\sum_{m=0}^{\infty }%
\frac{(-1)^{n+m}v^{n}\Delta ^{2m} t^{n+2m}}{4^{m}n!m!\Gamma \left(
\frac{n+2m+2}{2}\right) (n+2m+1)},  \label{needsanumber}
\end{equation}
which can also be obtained directly by expanding Eq.~(\ref{Stld2})
in $1/\alpha $ and then taking the inverse Laplace transform. The
double series can be rearranged to obtain the following expansion
in the detuning:

\begin{equation}
S(t)_{av}=\sqrt{\frac{vt}{2}}
\int_{0}^{\infty}\frac{dp}{\sqrt{p}}\,e^{-p^{2}}
\sum_{m=0}^{\infty} \frac{1}{m!} \left( -\frac{\Delta^{2}pt}{2v}
\right)^{m} J_{2m+1} \left( \sqrt{8vpt} \right) .
\label{deltaexpansion}
\end{equation}
The first term in this expansion, the on-resonance signal, can be
written as
\begin{eqnarray}
S(t)_{av}\biggr|_{\Delta =0} &=&\left[ \frac{\sqrt{\pi }}{2}vt\
_{0}F_{2}\left( 1,\frac{3}{2};\frac{v^{2}t^{2}}{4}\right) \right.
\nonumber
\\
&&\left. +\frac{1}{2}-\frac{1}{2}\ _{0}F_{2}\left( \frac{1}{2},\frac{1}{2};%
\frac{v^{2}t^{2}}{4}\right) \right] ,  \label{Sres}
\end{eqnarray}
which can also be obtained by setting $\Delta $ equal to zero in
Eq.~(\ref{hypergeo}) and summing the resulting series. Plots of
$S(t)_{av}$ for several values of $\Delta $ are shown
 in Fig.~\ref
{figure1}. Notice that the initial slope of the signal is independent of $%
\Delta .$ This feature translates into resonance widths that vary as $1/t$ for small $%
t $ - the so-called transform broadening discussed by Thomson
\emph{et al.}\cite {thomson,thomson2}. With $v$ on the order of
MHz, this initial rise occurs in a fraction of a $\mu $s. Another
point of interest is that the averaging is much more effective at
smoothing out the oscillations for $\Delta =0$ than it is for
$\Delta \neq 0.$

We can construct a measure of the width from the first two terms in the $%
\Delta $-expansion, Eq.~(\ref{deltaexpansion}). If
\begin{equation}
S(t)_{av}=S_{0}\left( t\right) +\Delta ^{2}S_{1}\left( t\right) +{\cal O}%
\left( \Delta ^{4}\right) ,
\end{equation}
and we define
\begin{equation}
    w = \sqrt{-\frac{S_{0}\left( t \right)}{S_{1}\left( t \right)}},  \label{width}
\end{equation}
then $2w$ would be the FWHM if the lineshape were Lorentzian. The
quantity $w/v$ is plotted in Fig.~\ref{figure2}. For small $vt$,
we have explicitly $w = \sqrt{12}/t$. While the $1/t$ behavior
follows from the transform broadening argument, or even simply
from dimensional analysis, the coefficient $\sqrt{12}$ is a
prediction of the detailed theory.

The lineshape at saturation ($t\rightarrow \infty $) can be
obtained using the relation in Eq.~(\ref{trik}), which yields
\begin{equation}
S(t)_{av}=\frac{1}{\sqrt{\pi }}\int_{0}^{\infty }dy~e^{-y^{2}}\frac{%
\left[ 1-\cos \left( \sqrt{\Delta ^{2}+\frac{v^{2}}{y^{2}}}
~t \right) \right] }{%
1+\frac{\Delta ^{2}}{v^{2}}y^{2}}. \label{TT}
\end{equation}
For large times the cosine term above averages to zero, giving
\begin{equation}
S(t\rightarrow \infty )_{av}=\frac{\sqrt{\pi }}{2}\frac{v}{\Delta}\exp \left( \frac{v^{2}%
}{\Delta ^{2}}\right) {\rm erfc}\left( \frac{v}{\Delta }\right) ,
\label{Sinf}
\end{equation}
which can also be extracted from the small $\alpha $ behavior of
Eq.~(\ref{Stld2}). This lineshape is plotted in
Fig.~\ref{figure3}. The FWHM is approximately $4.6v.$ Also plotted
is a Lorentzian lineshape with the same height and FWHM. Notice
that this lineshape is sharper than the Lorentzian for small
$\Delta $ and falls off more slowly for large $\Delta$.

\section{Sparse $ss^{\prime }\rightarrow pp^{\prime }$ in a bath of $%
sp\rightarrow ps$}

\label{decay}

We again consider one atom, initially in state $s^{\prime }$, in
interaction with a gas of $s$ atoms via the $ss^{\prime
}\rightarrow pp^{\prime }$ process, but we now allow for
$sp\rightarrow ps$ processes by modifying Eqs.~(1) to
\setcounter{tlet}{1}
\renewcommand{\theequation}{\arabic{equation}\alph{tlet}}
\begin{eqnarray}
\label{withu}
 i~\dot{a}_{0} &=&\Delta ~a_{0}+\sum_{k}V_{k}~c_{k}, \\
\stepcounter{tlet}\addtocounter{equation}{-1}\label{a0.}
 i~\dot{c}_{k} &=&V_{k}~a_{0}+\sum_{l}U_{kl}c_{l}\,.
\label{ck.U}
\end{eqnarray}
\renewcommand{\theequation}{\arabic{equation}}
where $U_{kl}=\mu ^{2}/r_{kl}^{3}$ and the other symbols have the
same meaning as in Eqs.~(1). Although Eqs.~(24) are exact, they
are difficult to solve exactly. We therefore introduce the
equation
\begin{equation}
i~\dot{c}_{k}=V_{k}~a_{0}-i\gamma c_{k}\,,  \label{ck.gam}
\end{equation}
in place of Eq.~(\ref{ck.U}), where $\gamma $ is an effective
inverse lifetime. In a crude way, the $-i\gamma c_{k}$ term in
Eq.~(\ref{ck.gam}) describes the ``walking away'' of the $p$
states from the neighborhood of the resonating atom at the origin.
One can see from Eq.~(\ref{ck.U}) that the model of
Section~\ref{dilute} is recovered in the limit $\mu \ll \mu
^{\prime }$.

When Eq.~(\ref{ckdot}) of the last section is replaced with
(\ref{ck.gam}),
the solution of the new system can be obtained by replacing $\Delta $ with $%
\Delta +i\gamma $ in Eq.~(\ref{ck}) and in the corresponding equation for $%
a_{0}(t).$ Then
\begin{eqnarray}
\,L(t) &\equiv &\sum_{k=1}^{N}\left| c_{k}\left( t\right) \right|
^{2} \nonumber \\ &=&\frac{2{\cal V}^{2}~e^{-\gamma t}~\left[
\cosh \left( yt\right) -\cos \left( xt\right) \right]
}{x^{2}+y^{2}},  \label{Sck2}
\end{eqnarray}
where $x$ and $y$ are real numbers that satisfy
\begin{equation}
\left( x+iy\right) ^{2}=(\Delta +i\gamma )^{2}+4{\cal V}^{2}.
\label{z}
\end{equation}

At resonance, we have
\begin{equation}
\left( x,y\right) =\left\{
\begin{array}{cl}
\left( \sqrt{4{\cal V}^{2}-\gamma ^{2}},0\right) & \mbox{if
$\gamma <2{\cal V}$,} \\ \left( 0,\sqrt{\gamma ^2 - 4{\cal
V}^{2}}\right) & \mbox{if $\gamma >2{\cal V}$.}
\end{array}
\right.
\end{equation}
Note that $L\left( t\right) $ decays more slowly in the latter
case. Hence
the large-$t$ limit of $L\left( t\right) _{av}$ will be dominated by small $%
{\cal V}$. This remains true off-resonance as well. The
small-${\cal V}$ expansion for $y$ is
\begin{equation}
y=\gamma -%
{\displaystyle {2{\cal V}^{2}\gamma  \over \Delta ^{2}+\gamma ^{2}}}%
\end{equation}
Thus, we can conclude that for the leading behavior at large
times,
\begin{equation}
\lim_{t\rightarrow \infty }\,L\left( t\right) \propto \exp \left[ -%
{\displaystyle {2{\cal V}^{2}\gamma  \over \Delta ^{2}+\gamma ^{2}}}%
t\right] .
\end{equation}
Using Eq.~(\ref{ave}), we obtain an asymptotic behavior of the
average
\begin{equation}
\lim_{t\rightarrow \infty }\,L\left( t\right) _{av}\propto \exp \left( -%
\sqrt{\gamma _{eq}t}\right) ,  \label{asymptotics}
\end{equation}
where
\begin{equation}
\gamma _{eq}=%
{\displaystyle {2v^{2}\gamma  \over \Delta ^{2}+\gamma ^{2}}}%
.  \label{gammaeq}
\end{equation}

Let us now examine the full time dependence. We begin with the
Laplace transform of \thinspace $L(t)$ which is
\begin{equation}
\tilde{L}(\alpha )=\frac{2s{\cal V}^{2}}{s^{4}+\left(
x^{2}-y^{2}\right) s^{2}-x^{2}y^{2}},  \label{Ltild}
\end{equation}
where $s=\alpha +\gamma $. Substituting $x$ and $y$ from
Eq.~(\ref{z}) gives
\begin{equation}
\tilde{L}(\alpha )=\frac{2s{\cal V}^{2}}{s^{4}+\left( \Delta
^{2}-\gamma ^{2}+4{\cal V}^{2}\right) s^{2}-\Delta ^{2}\gamma
^{2}}.  \label{Ltld}
\end{equation}

The signal is still proportional to $N^{\prime }S(t)$, with
$S(t)=1-\left| a_{0}(t)\right| ^{2}$. It is no longer true that
$1-\left| a_{0}\right| ^{2}=\sum_{k}\left| c_{k}\right| ^{2}$,
because $\sum_{k}\left| c_{k}\right| ^{2}$ is not the total
probability of finding a $p$ state, but only the probability of
finding a $p$ state that has not ``walked away''. However, the
relation
\begin{equation}
{\displaystyle {d \over dt}}%
\left( \left| a_{0}\right| ^{2}+\sum_{k}\left| c_{k}\right|
^{2}\right) =-2\gamma \sum_{k}\left| c_{k}\right| ^{2},
\end{equation}
with the usual initial conditions, enables us to obtain
\begin{equation}
S(t)=L(t)+2\gamma \int_{0}^{t}L(t^{\prime })\,dt^{\prime },
\end{equation}
and
\begin{equation}
\tilde{S}(\alpha )=\left( 1+%
{\displaystyle {2\gamma  \over \alpha }}%
\right) \tilde{L}(\alpha ).  \label{needsanametoo}
\end{equation}
For $\gamma >0$, $S(t)$ rises from 0 to 1 (or more accurately to $(1+$ $%
N^{\prime }/N)^{-1},$ as seen in the next section), while for
$\gamma =0$ it saturates at $2{\cal V}^{2}/\left( \Delta
^{2}+4{\cal V}^{2}\right) $. What
happens for small $\gamma $ is a quick rise (on the average time scale of $%
v^{-1}$ ) to $2{\cal V}^{2}/\left( \Delta ^{2}+4{\cal
V}^{2}\right) $, followed by a slow rise to 1.

Averaging these equations is done as in Section~\ref{dilute}. For
instance, averaging $L\left( \alpha \right) $ yields
\begin{equation}
\widetilde{L}\left( \alpha \right) _{av}=%
{\displaystyle {\sqrt{\pi } \over 2}}%
\frac{v}{sB}~\exp \left( \frac{v^{2}}{B^{2}}\right) {\rm erfc}\left( \frac{v%
}{B}\right) ,  \label{LAII}
\end{equation}
where
\begin{equation}
B^{2}=\left( s^{2}+\Delta ^{2}\right) \left( s^{2}-\gamma
^{2}\right) /s^{2},
\end{equation}
which is the analog of Eq.~(\ref{Stld2}). The inverse Laplace
transform of this expression is rather complicated, but it
simplifies immensely at resonance, becoming
\begin{equation}
L\left( t\right) _{av}=%
{\displaystyle {\sqrt{\pi } \over 2}}%
v\sum_{n=0}^{\infty }%
{\displaystyle {\left( -1\right) ^{n} \over n!}}%
\left(
{\displaystyle {2v^{2} \over \gamma }}%
\right) ^{n/2}e^{-\gamma t}\int_{0}^{t}d\tau \,
\tau^{n/2}I_{n/2}\left( \gamma \tau \right) .  \label{Iform}
\end{equation}
This result can also be expressed in terms of generalized
hypergeometrics, and in that form appears similar to
Eq.~(\ref{hypergeo}). One can use the asymptotic behavior of the
Bessel function in Eq.~(\ref{Iform}) to confirm the large-$t$
dependence argued earlier in Eq.~(\ref{asymptotics}). An expansion
in inverse powers of $\alpha $, leading to an expression similar
to Eq.~(\ref{needsanumber}), is easily obtained by computer.
Fig.~\ref{figure4} shows the average signal $S\left(t\right)_{av}$
for $\gamma =v,$ and $\Delta/v $ having the values $0,$ $1,$ and
$2$. One can see that the approach to one becomes slower as the
detuning is increased, as predicted by Eqs.~(\ref{asymptotics})
and (\ref{gammaeq}). This behavior is the same as that found by
Mourachko \emph{et al.}\cite{mourachko} under the assumption that
the time development of the $c_{k}$ is governed by a diffusion
equation. We see that in the present approach it arises simply
from phase averaging. The result for the on-resonance signal is
shown in Fig.~\ref {figure5}. As one would expect, for small
values of $\gamma $ the small-$t$ features of Fig.~\ref{figure5}
resemble those in Fig.~\ref{figure1} of Section~\ref{dilute},
where the $sp\rightarrow ps$ process is neglected entirely.

\section{Interacting resonant process}

\label{spin}

We now consider $N$ atoms at positions ${\mathbf{r}}_{k}$ which
are initially in state $s$ and $N^{\prime }$ atoms at positions
${\mathbf{r}}_{k^{\prime }}$ which are initially in state
$s^{\prime }$. We represent the $s$ and $p$ states at
${\mathbf{r}}_{k}$ with the down and up states of an effective
spin ${\mathbf{\sigma }}_{k},$ and similarly represent the
$s^{\prime }$ and $p^{\prime }$ states at ${\mathbf{r}}_{k^{\prime
}}$ with a spin ${\mathbf{\sigma}}_{k^{\prime }}$. The Hamiltonian
is then
\begin{eqnarray}
&&\sum_{k}\left[ \varepsilon _{s}+(\varepsilon _{p}-\varepsilon
_{s})\sigma _{k}^{+}\sigma _{k}^{-}\right] +\sum_{k^{\prime
}}\left[ \varepsilon _{s}^{\prime }+(\varepsilon _{p}^{\prime
}-\varepsilon _{s}^{\prime })\sigma _{k^{\prime }}^{+}\sigma
_{k^{\prime }}^{-}\right]  \nonumber \\ &&+\sum_{kk^{\prime
}}V_{kk^{\prime }}\left[ \sigma _{k}^{+}\sigma _{k^{\prime
}}^{+}+\sigma _{k}^{-}\sigma _{k^{\prime }}^{-}\right]
+\sum_{k,l\neq k}U_{kl}\sigma _{k}^{+}\sigma _{l}^{-},  \label{H}
\end{eqnarray}
where
\begin{equation}
V_{kk^{\prime }}=\frac{\mu \mu ^{\prime }}{\left| {\bf r}_{k}-{\bf r}%
_{k^{\prime }}\right| ^{3}}\ \ \ {\rm and}\ \ \ U_{kl}={\frac{\mu ^{2}}{%
\left| {\bf r}_{k}-{\bf r}_{l}\right| ^{3}}}.  \label{V}
\end{equation}
For the experiments under consideration $\mu\approx 4\mu ^{\prime
},$ and the $sp$-$ps$ coupling $\mu ^{2},$ which leads to
``spin diffusion'', is larger than the $ss^{\prime }$-$%
pp^{\prime }$ coupling $\mu\mu'$, which is responsible for the
resonant energy transfer.

The strategy is to write down the evolution equations for the spin
variables, take the expectation values over the initial state
$\left| i\right\rangle $ (which consists of all spins down),
discard the expectation values of quantities that fluctuate
incoherently, solve the resulting equations, and finally average
over the atomic coordinates ${\bf r}_{k}$ and ${\bf r}_{k^{\prime
}}$. The experimental signal is proportional to the number of
$p^{\prime }$ states, $\sum_{k^{\prime }}n_{k^{\prime }}$, where
\begin{equation}
n_{k^{\prime }}=%
{\displaystyle {1 \over 2}}%
\left( 1+\left\langle \sigma _{k^{\prime }}^{z}\right\rangle
\right) . \label{n'}
\end{equation}
It should be noted that $\left| i\right\rangle $ is not the ground
state, nor does the system relax to the ground state, because it
is essentially decoupled from other degrees of freedom for the
duration of the experiment.

We start with basic equations of motion such as
\begin{eqnarray}
i%
{\displaystyle {d \over dt}}%
\sigma _{k^{\prime }}^{+} &=&\Delta \sigma _{k^{\prime
}}^{+}+\sigma _{k^{\prime }}^{z}\sum_{k}V_{k^{\prime }k}\sigma
_{k}^{-},  \label{s'+} \\
i%
{\displaystyle {d \over dt}}%
\sigma _{k}^{+} &=&\sigma _{k}^{z}\sum_{k^{\prime }}V_{kk^{\prime
}}\sigma _{k^{\prime }}^{-}+\sigma _{k}^{z}\sum_{l\neq
k}U_{kl}\sigma _{l}^{+}, \label{s+} \\
i%
{\displaystyle {d \over dt}}%
\sigma _{k^{\prime }}^{z} &=&2\sum_{k}V_{k^{\prime }k}\left(
\sigma _{k}^{+}\sigma _{k^{\prime }}^{+}-\sigma _{k}^{-}\sigma
_{k^{\prime }}^{-}\right) ,  \label{sz}
\end{eqnarray}
where $\Delta $ is the detuning from resonance\cite{fnote1}, and
obtain
\begin{eqnarray}
{\displaystyle {1 \over 2}}%
{\displaystyle {d^{2} \over dt^{2}}}%
\sigma _{k^{\prime }}^{z} &=&-\sum_{kl^{\prime }}V_{k^{\prime
}k}V_{kl^{\prime }}\sigma _{k}^{z}\left( \sigma _{k^{\prime
}}^{-}\sigma _{l^{\prime }}^{+}+\sigma _{k^{\prime }}^{+}\sigma
_{l^{\prime }}^{-}\right) \nonumber \\ &&-\sum_{kl}V_{k^{\prime
}k}V_{lk^{\prime }}\sigma _{k^{\prime }}^{z}\left( \sigma
_{k}^{-}\sigma _{l}^{+}+\sigma _{k}^{+}\sigma _{l}^{-}\right)
\nonumber \\ &&-\sum_{k,l\neq k}V_{k^{\prime }k}U_{kl}\sigma
_{k}^{z}\left( \sigma _{k^{\prime }}^{+}\sigma _{l}^{+}+\sigma
_{k^{\prime }}^{-}\sigma _{l}^{-}\right)  \nonumber \\ &&-\Delta
\sum_{k}V_{k^{\prime }k}\left( \sigma _{k^{\prime }}^{+}\sigma
_{k}^{+}+\sigma _{k^{\prime }}^{-}\sigma _{k}^{-}\right) .
\label{s2'}
\end{eqnarray}
We reduce the expectation value of this equation as follows. In
the first line, we assume that $\left\langle \sigma _{k}^{z}\left(
\sigma _{k^{\prime }}^{-}\sigma _{l^{\prime }}^{+}+\sigma
_{k^{\prime }}^{+}\sigma _{l^{\prime }}^{-}\right) \right\rangle $
is negligible unless $k^{\prime }=l^{\prime },$ in which case it
reduces to $\left\langle \sigma _{k}^{z}\right\rangle .$
Similarly, $\left\langle \sigma _{k^{\prime }}^{z}\left( \sigma
_{k}^{-}\sigma _{l}^{+}+\sigma _{k}^{+}\sigma _{l}^{-}\right)
\right\rangle $ is replaced by $\left\langle \sigma _{k^{\prime
}}^{z}\right\rangle \delta _{kl}$. The third line contains only
incoherent terms, which we treat by assuming that they lead to a
Lorentzian broadening by letting
\begin{equation}
\sigma _{k}^{z}\sum_{l\neq k}U_{kl}\sigma _{l}^{+}\rightarrow
i\gamma \sigma _{k}^{+}  \label{Lor}
\end{equation}
when computing these terms. With this crude approximation and Eq.~(\ref{sz}%
), we obtain
\begin{equation}
\sum_{kl}V_{k^{\prime }k}U_{kl}\left\langle \sigma _{k}^{z}\left(
\sigma _{k^{\prime }}^{+}\sigma _{l}^{+}+\sigma _{k^{\prime
}}^{-}\sigma
_{l}^{-}\right) \right\rangle =%
{\displaystyle {\gamma  \over 2}}%
{\displaystyle {d \over dt}}%
\left\langle \sigma _{k^{\prime }}^{z}\right\rangle \,.
\end{equation}
Collecting results, we obtain
\begin{eqnarray}
{\displaystyle {1 \over 2}}%
{\displaystyle {d^{2} \over dt^{2}}}%
\left\langle \sigma _{k^{\prime }}^{z}\right\rangle
&=&-\sum_{k}V_{k^{\prime }k}V_{kk^{\prime }}\left( \left\langle
\sigma _{k}^{z}\right\rangle +\left\langle \sigma _{k^{\prime
}}^{z}\right\rangle \right)  \nonumber \\
&&-%
{\displaystyle {\gamma  \over 2}}%
{\displaystyle {d \over dt}}%
\left\langle \sigma _{k^{\prime }}^{z}\right\rangle -\Delta
g_{k^{\prime }}\,,  \label{aveq'}
\end{eqnarray}
where
\begin{equation}
g_{k^{\prime }}(t)=\sum_{k}V_{k^{\prime }k}\left\langle \sigma
_{k^{\prime }}^{+}\sigma _{k}^{+}+\sigma _{k^{\prime }}^{-}\sigma
_{k}^{-}\right\rangle .
\end{equation}
Using arguments similar to those used to derive Eq.~(\ref{aveq'}),
we obtain the approximate equation
\begin{equation}
{\displaystyle {dg_{k^{\prime }} \over dt}}%
=\Delta
{\displaystyle {dn_{k^{\prime }} \over dt}}%
-\gamma\left( g_{k^{\prime }}-\bar{g} \right) ,  \label{dg}
\end{equation}
where $\bar{g}=$ $%
{\displaystyle {1 \over N^{\prime }}}%
\sum_{k^{\prime }}g_{k^{\prime }}$ and the $\gamma \overline{g}$
term has been put in by hand to preserve conservation. These
equations imply
\begin{equation}
{\displaystyle {d\bar{g} \over dt}}%
=\Delta
{\displaystyle {dn^{\prime } \over dt}}%
\,,  \label{dgbar}
\end{equation}
where
\begin{equation}
n^{\prime }=%
{\displaystyle {1 \over N^{\prime }}}%
\sum_{k^{\prime }}n_{k^{\prime }}\,,  \label{nprime}
\end{equation}
with the initial condition $g_{k^{\prime }}(0)=0.$

At this point, it is more convenient to work with the $p^{\prime
}$ occupation $n_{k^{\prime }},$ defined in Eq.~(\ref{n'}), and
with the corresponding $p$ occupation, $n_{k}$. For comparison
with experiment we will need the sample average, which corresponds
to $S\left( t \right)$ of Sections~\ref{dilute} and \ref{decay}.
Taking Laplace transforms of Eqs.~(\ref{aveq'}) and (\ref{dg}), we
obtain
\begin{eqnarray}
&&\left( \alpha ^{2}+2\sum_{k}V_{kk^{\prime }}^{2}+\gamma \alpha +%
{\displaystyle {\alpha \Delta ^{2} \over \alpha +\gamma }}%
\right) \tilde{n}_{k^{\prime }}  \nonumber \\
&=&%
{\displaystyle {2 \over \alpha }}%
\sum_{k}V_{kk^{\prime }}^{2}-2\sum_{k}V_{kk^{\prime }}^{2}\tilde{n}%
_{k}-\Delta ^{2}%
{\displaystyle {\gamma  \over \alpha +\gamma }}%
\tilde{n}^{\prime },  \label{ntild'}
\end{eqnarray}
where the tildes denote Laplace transforms and $\tilde{n}^{\prime
}(\alpha )$ is the sample average of $\tilde{n}_{k^{\prime
}}(\alpha ).$

We can obtain a similar equation for $\tilde{n}_{k}$, and we are
then left with coupled equations which we are not able to solve.
However, if $N\approx N'$
we can simplify the problem by assuming that the strong $sp$-$%
ps$ coupling acts to randomize the $\sigma _{k}$ spins, while the
resonant $ss^{\prime }$-$pp^{\prime }$ processes are entirely
coherent. Accordingly, in Eq.~(\ref{ntild'}) we replace
$\tilde{n}_{k}$ with its sample average $\tilde{n}$ and obtain an
equation for $\tilde{n}_{k^\prime }$ alone by using the exact
relation $Nn=N^{\prime }n^{\prime },$ which follows
from the fact that $p$ and $p^{\prime }$ states are created in pairs by the $%
V_{kk^{\prime }}$ term in Eq.~(\ref{H}). We solve for
$\tilde{n}_{k^{\prime }}$ and average over a random distribution
of atoms, replacing the sample average $(1/N^{\prime
})\sum_{k^{\prime }}n_{k^{\prime }}$ with an ensemble average, and
obtain
\begin{equation}
\tilde{n}^{\prime }= \frac{\left( 1/\alpha \right) \left( 1-C^{2}\tilde{F}%
_{av}\right)} {1+\frac{N^{\prime}}{N} -\left(\frac{N^{\prime
}}{N}C^{2}- \frac{\gamma \Delta ^{2}} {\alpha +\gamma }\right)
\tilde{F}_{av}} , \label{n'fin}
\end{equation}
where
\begin{equation}
C^{2}=\alpha ^{2}+\alpha \gamma +%
{\displaystyle {\alpha \Delta ^{2} \over \alpha +\gamma }}%
,  \label{AF}
\end{equation}
and $\tilde{F}_{av}$ is the ensemble average of
\begin{equation}
\tilde{F}=\frac{1}{C^{2}+2\sum_{k}V_{k}^{2}}\,.  \label{Fav}
\end{equation}
For $N=N^{\prime }$, Eq.~(\ref{n'fin}) simplifies to
\begin{equation}
\tilde{n}^{\prime }= \frac{\left( 1/\alpha \right) \left( 1-C^{2}\tilde{F}%
_{av}\right)} {2-\left(C^{2}- \frac{\gamma \Delta ^{2}} {\alpha +\gamma }%
\right) \tilde{F}_{av}} ,  \label{FavN=N'}
\end{equation}

$\tilde{F}_{av}$ is computed in the same way as the averages of
$\tilde{S}$ in Section~\ref{dilute} and $\tilde{L}$ in
Section~\ref{decay}. The analog of Eq.~(\ref{Stld2}) is
\begin{equation}
C^{2}\tilde{F}(\alpha )_{av}=1-\sqrt{%
{\displaystyle {\pi  \over 2}}%
}%
{\displaystyle {v \over C}}%
\exp \left( \frac{v^{2}}{2C^{2}}\right) {\rm erfc}\left( \frac{v}{\sqrt{2}C}%
\right) ,
\end{equation}
with $v$ given by Eq.~(\ref{v}) and $C$ by Eq.~(\ref{AF}). One way
to compute $n^{\prime }(t)$ is to expand $\tilde{n}^{\prime
}(\alpha )$, Eq.~(\ref{n'fin}), in inverse powers of $\alpha $.
This gives the analog of Eq.~(\ref{needsanumber}), but we have not
found a simple expression for the coefficients.

Fig.~\ref{figure6} shows the signal as calculated in
Eq.~(\ref{FavN=N'}) for $\gamma =v$, and $\Delta/v =0$, $1$, and
$2$. We see that the on-resonance signal is devoid of any
oscillations. Off-resonance, the oscillations are not completely
washed out by the phase averaging, and some
nonmonotonicity is evident. Consequently, the dip in the width seen in Fig.~%
\ref{figure2} is also present in this calculation, provided
$\gamma $ is not too large.

Fig.~\ref{figure6} differs from the corresponding
Fig.~\ref{figure4} of Section~\ref{decay} in one important
respect: the on-resonance signal saturates to $1/2$ rather than to
$1$. For general values of $N$ and $N'$, Eq.~(\ref{n'fin})
predicts that as $t\rightarrow\infty $ (i.e., as
$\alpha\rightarrow 0$) $N'n'$ approaches $NN'/\left( N+N'
\right)$, as expected from simple kinetics for the two-body
reaction $ss'\rightarrow pp'$.

As in the previous sections, the energy scale is set by $v$, which
according to Eq.~(\ref{v}) has the value $5.72\,\mu\mu'N/\Omega$.
As discussed by Anderson \emph{et al.}\cite{anderson}, this
quantity is of the correct order of magnitude to account for the
data.

\subsection{Reduction to the sparse limit}

We now show explicitly how the general spin-variable formalism
relates to the coefficients $a_{0}$ and $c_{k}$ of
Sections~\ref{dilute} and \ref{decay} when there is only a single
atom that can be in the $s^{\prime }$, $p^{\prime }$ pair of
states (i.e., when only a single spin variable of the type $\sigma
_{k^{\prime }}$ is present.) The key to the correspondence is that
in this case the initial state $\left| i\right\rangle $ (which
consists of all spins down) evolves to
\begin{equation}
\left| t\right\rangle =\left( a_{0}(t)+\sum_{q}c_{q}(t)\sigma
_{k^{\prime }}^{+}\sigma _{q}^{+}\right) \left| i\right\rangle
\end{equation}
at time $t$, where the spin variables now denote time-independent
(Schr\"{o}dinger) operators. With the understanding that ${\bf
r}_{k^{\prime }}=0$, we have:

\begin{equation}
\left\langle \sigma _{k^{\prime }}^{z}\right\rangle \equiv
\left\langle t\right| \sigma _{k^{\prime }}^{z}\left|
t\right\rangle =1-2\left| a_{0}(t)\right| ^{2},
\end{equation}
\begin{equation}
\left\langle \sigma _{k}^{z}\right\rangle =2\left| c_{k}(t)\right|
^{2}-1,
\end{equation}
\begin{equation}
\left\langle \sigma _{k^{\prime }}^{+}\sigma _{k}^{+}\right\rangle
=a_{0}(t)c_{k}^{\dagger }(t).
\end{equation}

It is then clear that the expectation value of Eq.~(\ref{sz}) is
equivalent to the following, which is itself a consequence of
Eqs.~(1):
\begin{equation}
{\displaystyle {d \over dt}}%
\left| a_{0}(t)\right| ^{2}=\sum_{k}V_{k}\left(
a_{0}c_{k}^{\dagger }-a_{0}^{\dagger }c_{k}\right) ,
\end{equation}
where $V_{k}$ is short for $V_{k^{\prime }k}$. Similarly, the equations for $%
\left\langle \sigma _{k}^{z}\right\rangle $ and $\left\langle
\sigma _{k^{\prime }}^{+}\sigma _{k}^{+}\right\rangle $ are
equivalent to the equations for $\left| c_{k}(t)\right| ^{2}$ and
$a_{0}(t)c_{k}(t)^{\dagger }$ that follow from Eqs.~(\ref{ck.U}).

When the Lorentzian approximation is made in Eq.~(\ref{s2'}) or
its analog for $\sigma _{k}^{z}$, it is possible to obtain a
closed set of equations for $\left\langle \sigma
_{k}^{z}\right\rangle $ and $\left\langle \sigma _{k^{\prime
}}^{z}\right\rangle $ that give the results reported in
Sections~\ref{dilute} and \ref{decay} for a sparse system. As
indicated by the Laplace transforms in Eqs.~(\ref{Ltld}) and
(\ref{needsanametoo}), the
equation for $L(t)=%
{\textstyle {1 \over 2}}%
\left( 1+\sum_{k}\left\langle \sigma _{k}^{z}\right\rangle \right)
$ must be
of the fourth order, and that for $S(t)=%
{\textstyle {1 \over 2}}%
\left( 1-\sum_{k}\left\langle \sigma _{k}^{z}\right\rangle \right)
$ of fifth order.

If Eq.~(\ref{n'fin}) were exact, it would correctly give the
signal in the limit $N^{\prime }\ll N$, and $n^{\prime }(t)$ would
be equal to $S(t)_{av}$ of Section~\ref{decay} if the values for
$\gamma $ used in the two models were assumed to be equal. In
addition, $n^{\prime }(t)$ would reduce to the exact $S(t)_{av}$
of Section~\ref{dilute} if $\gamma $ were taken to be zero. Since
the result for $n^{\prime }(t)$ is based on approximations that
are certainly not valid for $N^{\prime }\ll N$, it is already
comforting that, in that limit, it shows a general resemblance to
the correct behavior. We are currently working on better
approximations to apply to the spin-variable formalism.

\section{\protect\smallskip Discussion}

The system we are studying can exhibit a wide range of behaviors,
depending on the values of the parameters $\mu /\mu^{\prime }$ and
$N/N^{\prime }.$ At the same time, there are features that persist
(at least qualitatively) throughout the range of parameters.

\begin{figure}[tbp]
\begin{center}
\leavevmode \hbox{\epsffile{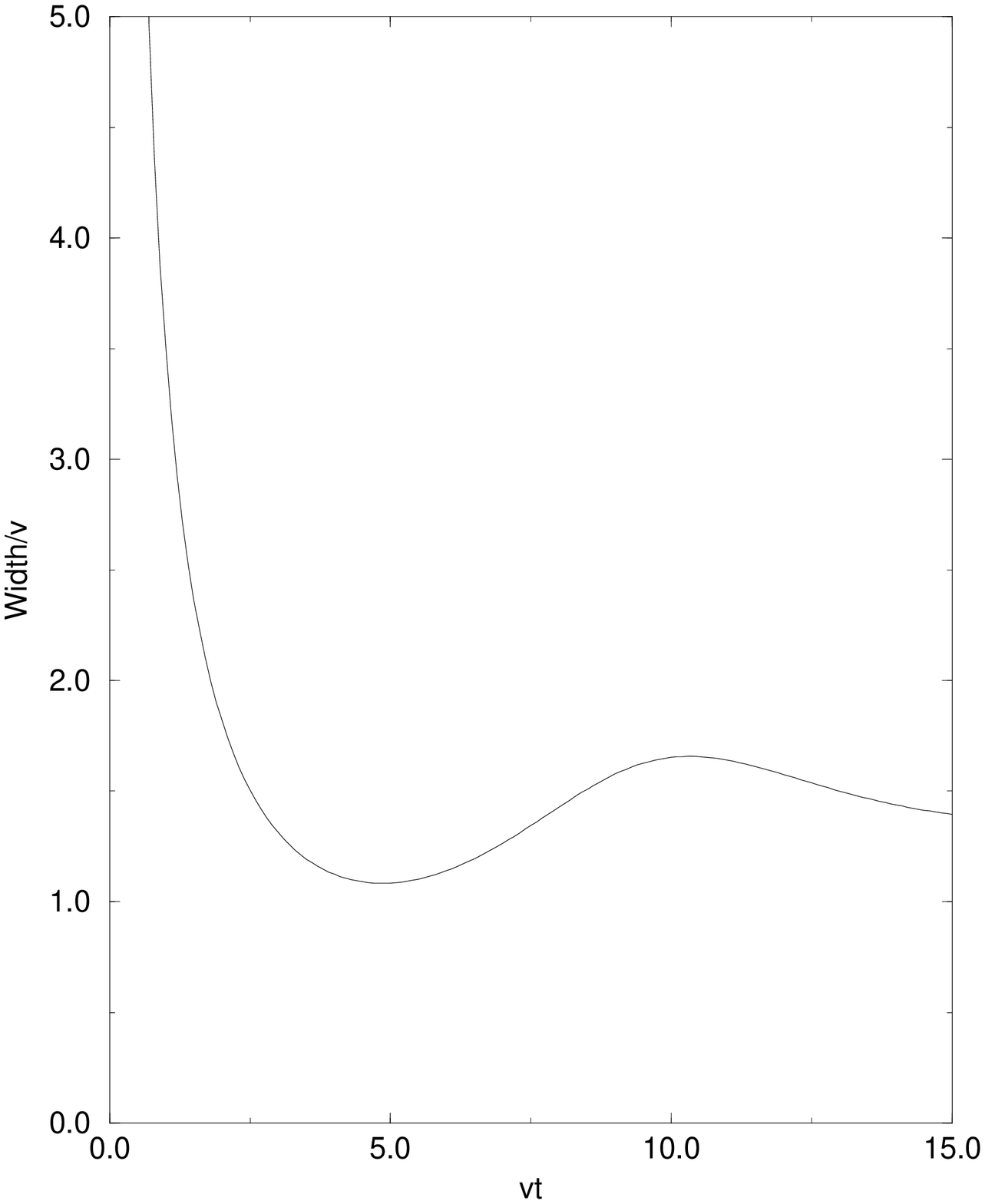}}
\end{center}
\caption{ The ``width" versus time, as given by Eq.~(\ref{width}),
shows a dip which is associated with the stronger and higher
frequency oscillations seen at larger values of $\Delta$ as shown
in Fig.~\ref{figure1}. } \label{figure2}
\end{figure}

\begin{figure}[tbp]
\begin{center}
\leavevmode \hbox{\epsffile{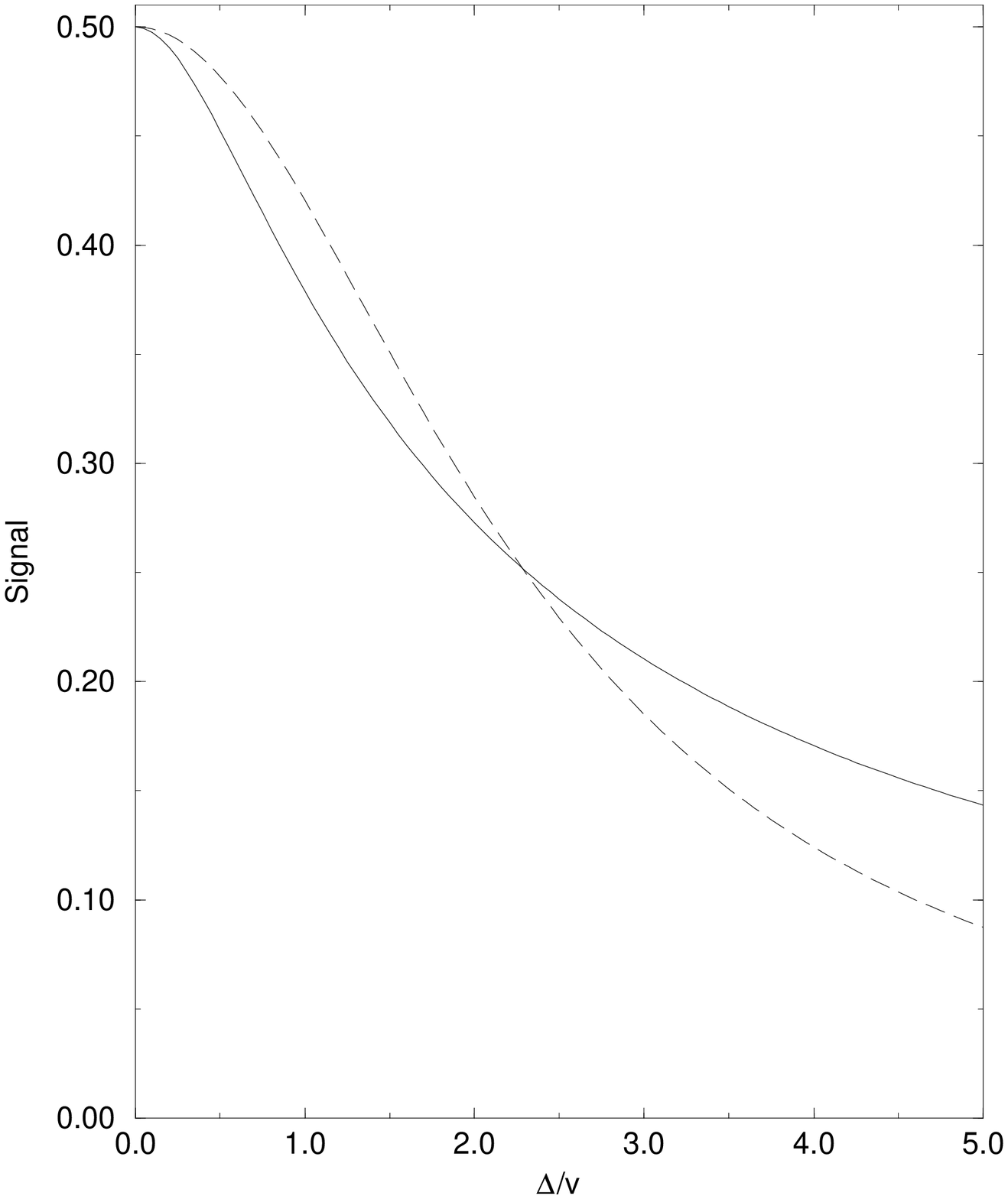}}
\end{center}
\caption{ The saturation ($t\rightarrow\infty$) lineshape,
Eq.~(\ref{Sinf}). It is not a Lorentzian but a Lorentzian averaged
as in Eq.~(\ref{TT}). The dashed line is a Lorentzian with the
same height and FWHM, drawn for comparison. } \label{figure3}
\end{figure}

\begin{figure}[tbp]
\begin{center}
\leavevmode \hbox{\epsffile{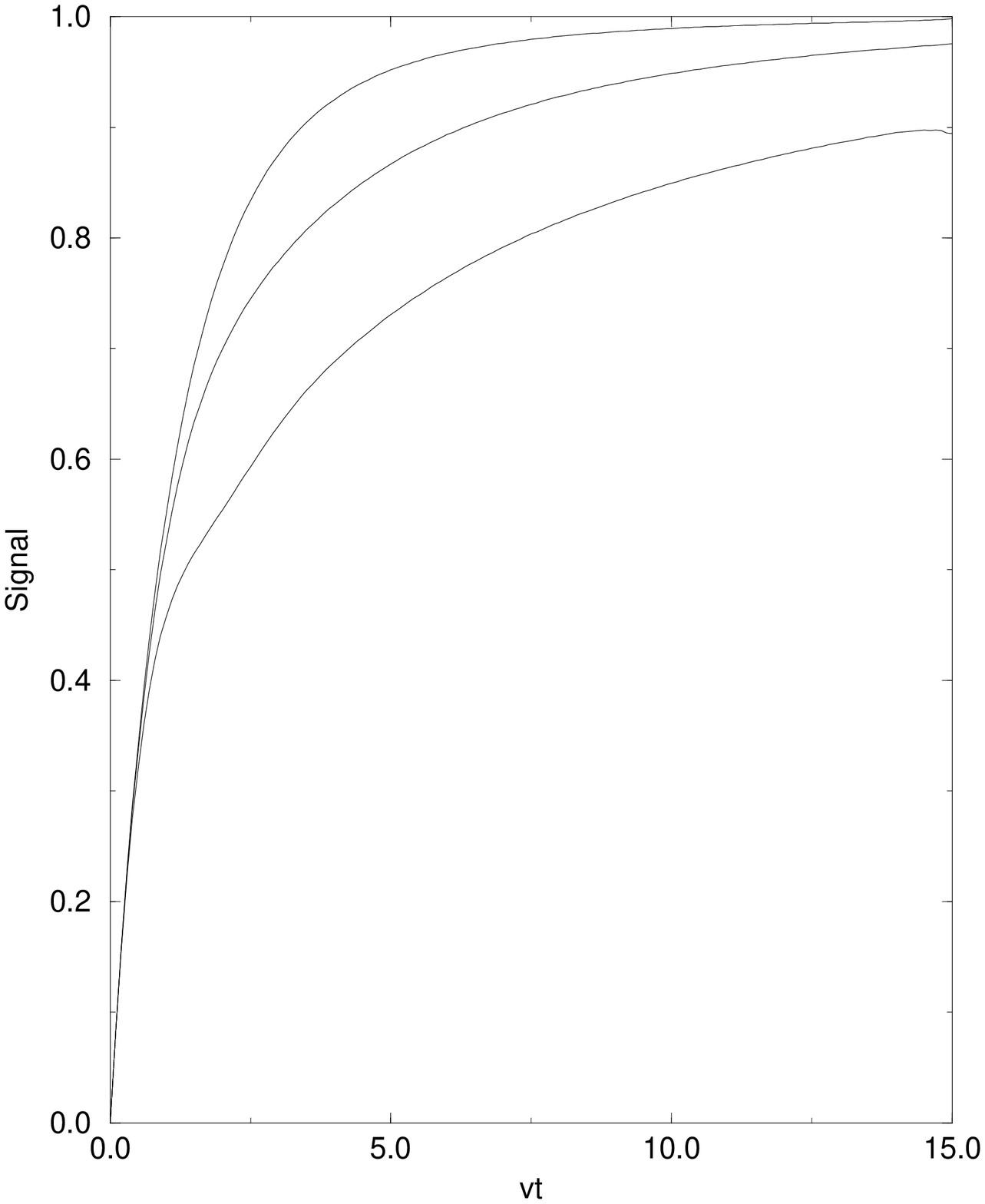}}
\end{center}
\caption{ The averaged signal, $S(t)_{av}$, for $\gamma=v$, and
$\Delta/v=0$, $1$, and $2$ (the upper, middle, and lower curves,
respectively) calculated from $\tilde{S}\left( \alpha
\right)_{av}=\left( 1+\frac{2\gamma}{\alpha} \right)
\tilde{L}\left( \alpha \right)_{av}$ and Eq.~(\ref{LAII}). The
approach to saturation goes like $\exp \left(
-\sqrt{\gamma_{eq}t}\right) $, with $\gamma_{eq}$ given by
Eq.~(\ref{gammaeq}). } \label{figure4}
\end{figure}

\begin{figure}[tbp]
\begin{center}
\leavevmode \hbox{\epsffile{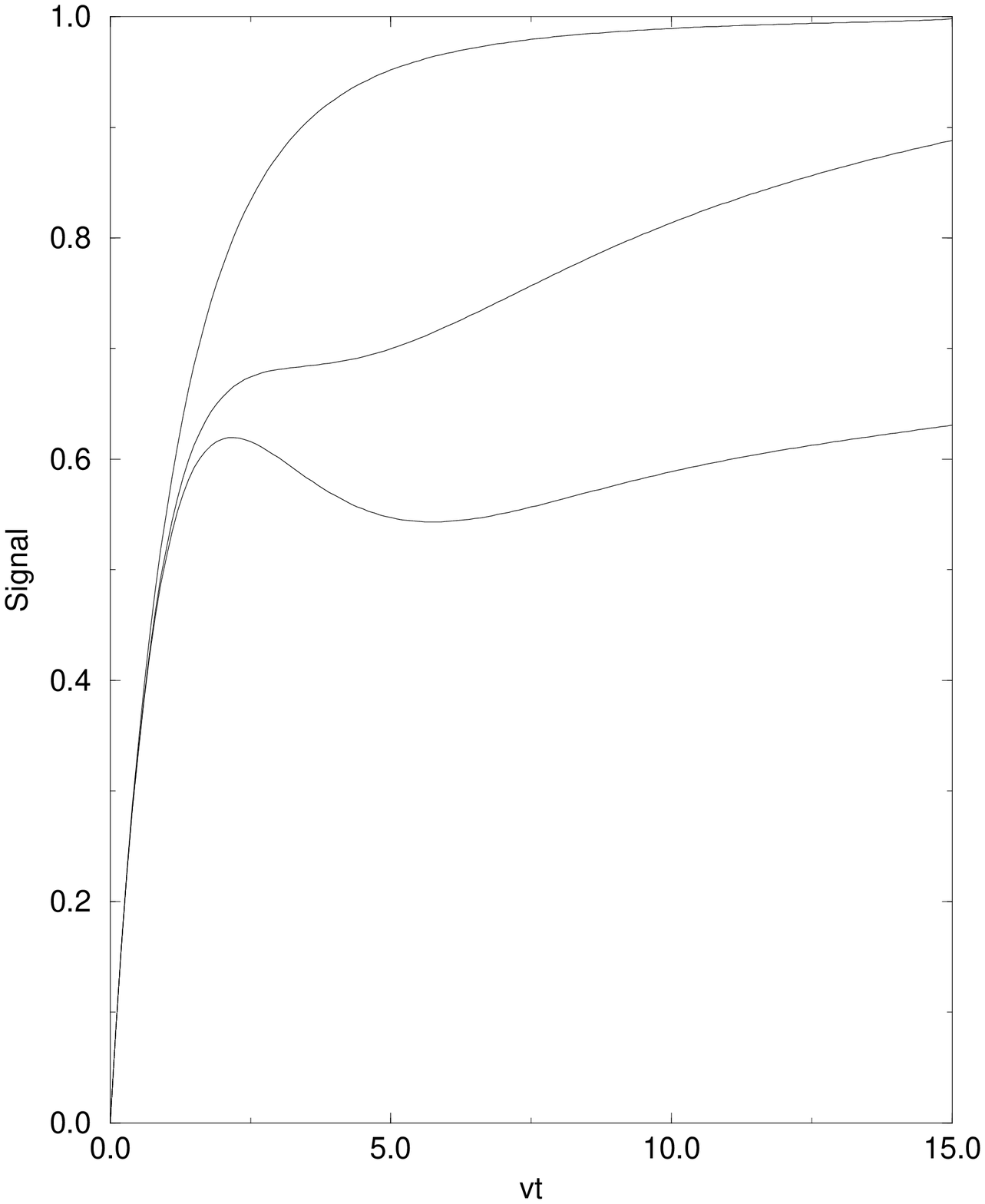}}
\end{center}
\caption{ $S\left( t \right)_{av}$, as in Fig.~\ref{figure4},
calculated at resonance $\left( \Delta = 0 \right)$, for inverse
lifetimes $\gamma/v=1$, $0.1$, and $0.02 $ (the upper, middle, and
lower curves, respectively). For the largest value of $\gamma$ the
oscillation is completely gone.} \label{figure5}
\end{figure}

\begin{figure}[tbp]
\begin{center}
\leavevmode \hbox{\epsffile{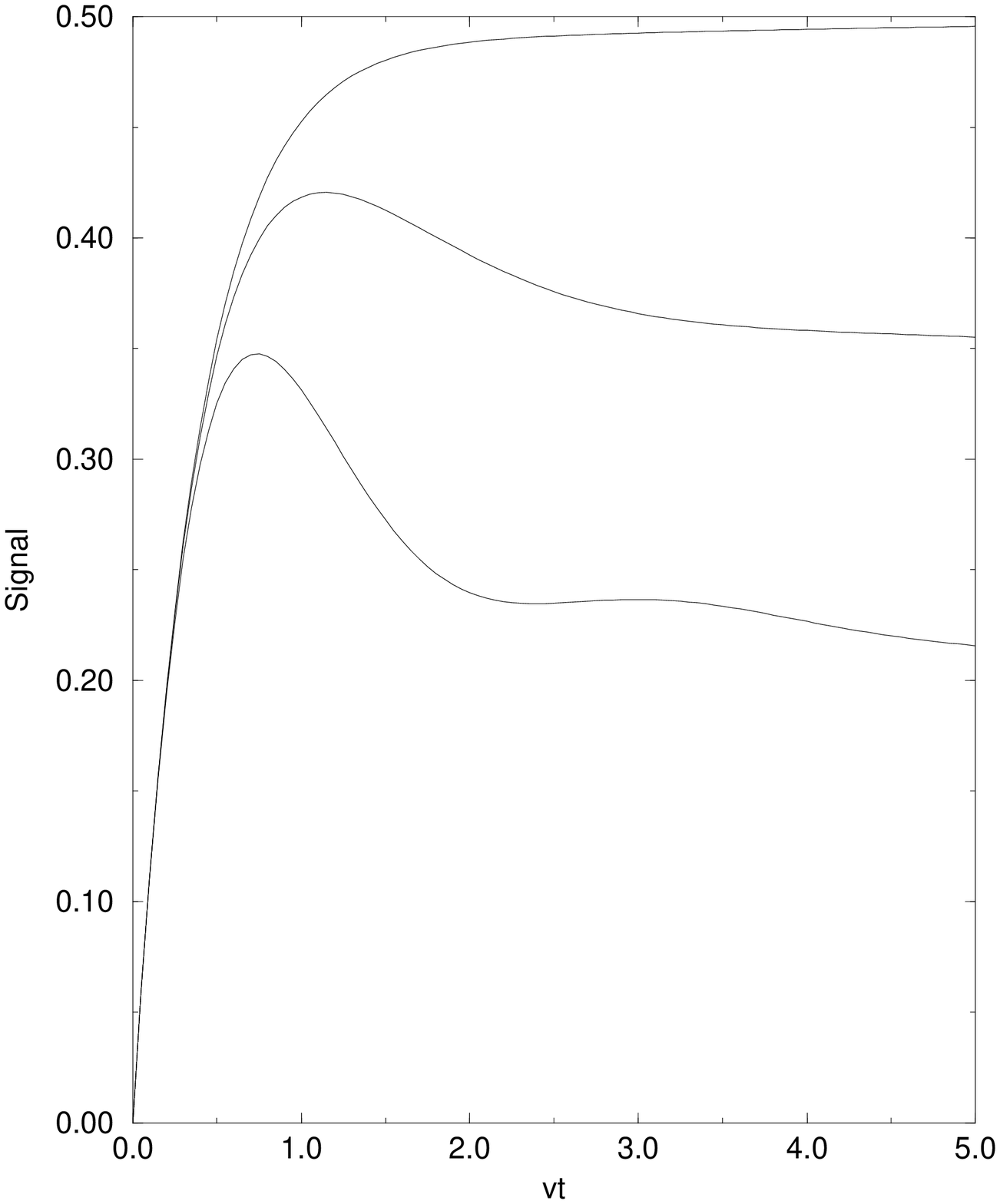}}
\end{center}
\caption{ The averaged signal, $n'(t)$, for $N=N'$, $\gamma=v$,
and $\Delta/v=0$, $1$, and $2$ (the upper, middle, and lower
curves, respectively), as calculated from Eq.~(\ref{FavN=N'}).
While the oscillations are ``damped out" for $\Delta=0$, they
remain for $\Delta/v=1$ and $2$. } \label{figure6}
\end{figure}

In Section~\ref{dilute}, we determine the exact behavior of a
frozen system consisting of a single $s^{\prime }$ state
interacting resonantly with a sea of $s$ states. That is to say,
we treat exactly the limit $\mu \ll \mu ^{\prime }$ and $N^{\prime
}\ll N$. A single $ss^{\prime }$ pair produces, of course, a
signal with a sinusoidal behavior, and the same is true, less
obviously, for a single $s^{\prime }$ interacting with a sea of
$s$ that are not mutually interacting. The phase interference
brought in by averaging over configurations causes the signal from
a sparse system of $s^{\prime }$ to take on the form shown in
Fig.~\ref{figure1}. Because the averaging does not completely wash
out the oscillations at small times, especially for nonzero values
of the detuning, the resonance linewidth has the nonmonotonic
behavior seen in Fig.~\ref{figure2}. The initial rise of the
averaged signal is linear in time, even though the unaveraged
signal is quadratic, and is given by $\sqrt{\pi}vt/2$, with $v$
defined in Eq.~(\ref{v}). The linewidth at large times is
approximately $4.6\,v$\cite{fnote2}. The fact that $v$ is sizably
larger than the simple estimate $\mu\mu ^{\prime }\left( N/\Omega
\right) $ reflects quantitatively the fact that ``close pairs" of
atoms are more heavily weighted in the average.

In Section~\ref{decay} we try to allow for any ratio $\mu /\mu
^{\prime }$, but we still keep $N^{\prime }\ll N$: physically, we
consider a sparse distribution of (initially) $s^{\prime }$ states
in a gas where $sp$-$ps$ flips take place
at a non-negligible rate. The first question we should answer is: after the $%
ss^{\prime }-pp^{\prime }$ transition, does the $p$ state ``walk
away'' to infinity or remain localized? In this paper, we have
assumed that localization does not take place, or, if it does, is
important only in the very sparse limit where the localization
length is smaller than $\left( \Omega /N\right) ^{1/3}$. Leaving
much room for future work, we have simply introduced a lifetime
$\gamma^{-1} $ for the $p$ state to ``stick around''. The spectral
function for the propagating $sp$-$ps$ flip is then simply a
Lorentzian. We have obtained tentative estimates of $\gamma $ when
it is entirely due to the $sp$-$ps$ flips, and we plan to report
these results in the near future. There can also be other
contributions to $\gamma $, arising for instance from
translational motion of the atoms. (The gas can be treated as
frozen to
a good approximation when the typical collision time $\tau $ exceeds $v^{-1}$%
, and certainly $\tau^{-1}$ contributes to $\gamma$.) In this
paper, we in effect add a damping term to the equations of
Section~\ref{dilute}, making the problem soluble again. The effect
of $\gamma $ is to further smooth out and wash away the
oscillatory behavior of the unaveraged signal. For large $\gamma $
all traces of the oscillations are wiped out and the signal
saturates according to $\exp \left( -\sqrt{\gamma _{eq}t}\right)
$, where $\gamma _{eq}$ is given by Eq.~(\ref{gammaeq}). The
saturated
linewidth is infinite, because the $%
s^{\prime }$ state eventually decays for all values of $\Delta $.
However, the signal at finite $t$ varies significantly with
$\Delta ,$ indicating that the linewidth has a minimum at
intermediate times. For small $t$ the signal is always independent
of $\Delta ,$ consistent with a linewidth proportional to $t^{-1}
$, as seen already in Fig.~\ref{figure2}.

In Section~\ref{spin} we introduce an effective spin Hamiltonian
to describe the resonating frozen gas for all values of the
parameters $\mu /\mu'$ and $N/N'$, and we discuss in particular
the case $N\approx N'$, which is relevant to the experiments
carried out so far. The only simplification we make in this spin
Hamiltonian is to neglect the potential responsible for the $%
s^{\prime }p^{\prime }$-$p^{\prime }s^{\prime }$ interaction. This
coupling can easily be included, although the resulting equations
of motion contain more terms and would be more difficult to
handle; it is in fact small for the experiments of Anderson
\emph{et al.}\cite{anderson,lowell}. Other resonant systems can be
described by similar spin Hamiltonians. The experiment of
Mourachko \emph{et al.}\cite{mourachko}, for example, consists of
a $pp-ss^{\prime } $ resonance in the presence of $sp$-$ps$ and
$s^{\prime }p-ps^{\prime }$ flips. A spin-one Hamiltonian
therefore describes this system, with the $-1,$ $0,$ and $1$ spin
states corresponding to the $s,$ $p,$ and $s^{\prime }$ atomic
states, respectively. In this case, there is only one set of spins
and all spins are initially in the $0$ state. The spin model we
have discussed in this paper consists of two sets of spins, which
complicates matters but at the same time allows one to consider
the limit in which one set is sparse.

There is practically no end to the variety of phenomena that this
type of spin Hamiltonian can describe, combining features from
many branches of physics. Atomic physics, random systems, spin
systems and magnetic resonance, and many-body theory are all
represented. The equilibrium behavior poses interesting problems
of statistical mechanics, but we are interested in the quantum
dynamics, starting from a prepared state. We have obtained and
solved approximate dynamical equations for the case $N\simeq
N^{\prime }$, using the Lorentzian approximation as in
Section~\ref{decay} to model the $sp$-$ps$ interaction. Much more
work can be done and is now in progress, and some of the results
may change when better approximations are introduced. In
particular, in a better theory the linewidth is likely to depend
on the $sp$-$ps$ interaction as well as on the $ss^{\prime
}$-$pp^{\prime }$. Nevertheless, the qualitative behavior of the
signal is common to all approximations we have tested so far and
matches what is seen in the experiments that initially motivated
this work. The important features that we already discussed are
all there: the sharp initial rise, the slow saturation, the
dependence on $\Delta $ that is less pronounced at small times and
corresponds to a time-dependent linewidth.

\label{discuss}

\acknowledgements We wish to thank T. Gallagher and J. Lowell for
many useful discussions. TB acknowledges the support of the
National Science Foundation under grant DMR9312476.

\appendix

\section{Angular averaging}

\label{angular-appendix} In Section~\ref{dilute} we averaged the quantity $%
e^{-\beta {\cal V}^{2}}$, where ${\cal V}^{2}=\sum_{k=1}^{N}V_{k}^{2}$ and $%
V_{k}=\mu \mu ^{\prime }/r_{k}^{3}$. In this appendix we include
the angular dependence of the dipole-dipole interaction assuming
that all of the dipoles point in the same direction, i.e. we
replace $V_{k}$ with
\begin{equation}
V(r,\theta )=-\frac{\mu _{1}\mu _{2}}{r^{3}}\left( 3\cos
^{2}\theta -1\right) .
\end{equation}
The analog of Eq.~(\ref{aveintsimple}) now has a nontrivial
angular part; it becomes
\begin{eqnarray}
&&2\pi \int_{0}^{\infty }r^{2}dr\int_{-\pi /2}^{\pi /2}\sin \theta
d\theta \left[ 1-e^{-\beta V^{2}(r,\theta )}\right]  \nonumber \\
&=&\frac{16\pi ^{3/2}}{9\sqrt{3}}~\mu \mu ^{\prime }~\sqrt{\beta
}.
\end{eqnarray}
The additional angular dependence changes $v$ by a factor
$4/3\sqrt{3}$.



\widetext


\begin{references}
\bibitem{fnote1} To obtain these expressions we take the equations of
motion that come about naturally from the Hamiltonian in
Eq.~(\ref{H}) and introduce a phase $\exp \left[ -i\left( \epsilon
_{p}-\epsilon _{s}\right) t\right] $ into the definitions of
$\sigma _{k}^{+}$ and $\sigma _{k^{\prime }}^{+}$.
\bibitem{fnote2} All widths discussed in this paper are in energy units.
There is an additional factor of $1/2\pi$ when computing the width
in frequency units.
\end{references}
\end{document}